\documentstyle[12pt]{article}
\begin{document}

\def\doublespace{\baselineskip = \normalbaselineskip \multiply\baselineskip
by 2}
\def\refpar{\hangindent 2truecm}

\def\ddd{\raise 6pt\hbox{\hskip .3pt {.}\kern -.8pt{.}\kern -.8pt{.}}}
\def\r{\rlap}

\def\mb#1{\setbox0=\hbox{#1}\kern-.025em\copy0\kern-\wd0
\kern-0.05em\copy0\kern-\wd0\kern-.025em\raise.0233em\box0}

\def\pmb#1{\setbox0=\hbox{$#1$}\kern-.025em\copy0\kern-\wd0
\kern-0.05em\copy0\kern-\wd0\kern-.025em\raise.0233em\box0}

\def\bfa{{\bf a}}
\def\bfu{{\bf u}}
\def\bfx{{\bf x}}
\def\cll{{\cal L}}
\def\parp{{\parallel}}
\def\pde{{\partial}}
\def\cali{{\cal I}}
\def\tr{{\rm tr}}
\def\sign{{\rm sign}}
\def\RR{{{I\negthinspace\!R}}}
\def\U{\Upsilon}
\def\O{\overline}
\def\V{\varphi}
\def\E{\varepsilon}
\def\P{\partial_}
\def\T{\vartheta}
\def\X{\xi}
\def\A{{\cal A}}
\def\B{{\cal B}}
\def\C{{\cal C}}
\def\hlf{{1\over2}}

\def\bfxi{{\pmb{\xi}}}
\def\bfp{{\pmb{\varphi}}}
\def\bfnab{{\pmb{\nabla}}}

%
%

\begin{center}
{\Large \bf Advection of vector fields by chaotic
flows}\footnote{Submitted for publication in the {\sl Proceedings of the
8th Florida
Workshop in Nonlinear Astronomy: Noise},
New York Academy of Sciences (1993).}

\end{center}
\vskip .5truein
\centerline{N.J. BALMFORTH,$^a$ P. CVITANOVI\'C,$^b$
G.R. IERLEY,$^c$ E.A. SPIEGEL$^a$ AND
G. VATTAY$^b$}
\bigskip
\centerline{\sl $^a$Astronomy Department}
\centerline{\sl Columbia University, New York, NY 10027}
\medskip
\smallskip
\centerline{\sl $^b$Niels Bohr Institute}
\centerline{\sl Blegdamsvej 17, DK-2100 Copenhagen}
\medskip
\smallskip
\centerline{\sl $^c$Scripps Institute of Oceanography}
\centerline{\sl UCSD, San Diego, CA 92037}
\bigskip
\bigskip
{\begin{center}
{\it ``The high average vorticity that is known to exist in
turbulent motion is caused by the extension of vortex filaments in an
eddying fluid.''}\\
{--- G.I. Taylor (1938)}
\end{center}

\vspace{10cm}

\pagebreak

\centerline{\bf THE PROBLEM}
\medskip

When the particles of a fluid are endowed with some
scalar density $S$, the density evolves
in time according to
$$
\partial_t S + {\bf u\cdot\nabla}S \equiv  {DS\over Dt}
= {\rm thermal\ noise.}
\eqno(1)
$$
The right-hand side represents the
microscopic spreading of $S$ on the molecular level,  and
can be thought of as noise added onto the fluid velocity ${\bf u}$.
It is normally described by a term like $\kappa{\bfnab}^2 S$
where $\kappa$ is a diffusivity. The
study of (1), especially for chaotic flows and turbulent
flows, has been extensively carried on for many decades.$^{1,2}$

Fluid motions also move vector fields around. An evolving
vector field ${\bf V}$ is governed by an equation of the form
$$
\partial_t {\bf V} + {\bf u\cdot\nabla}{\bf V} -
{\bf V\cdot\nabla u} \equiv {{\cal D}{\bf V}\over {\cal D}t}
= {\rm thermal\ noise.}
\eqno(2)
$$
The advective derivative in (1) is replaced by
a Lie derivative
${\cal D}{\bf V} / {\cal D}t$
in the evolution operator.
The extra term enters because the
evolution of a vector field involves not only being carried about by a
fluid particle but also being turned and stretched by the
motion of neighboring particles.  The thermal noise on the right
side is usually a simple diffusion term, at least for
the case of constant density. Density variations bring in some
unappetizing complications that we shall ignore.

If the right sides of (1) and (2) are zero (perfect fluid motion)
then $S$ and ${\bf V}$ are frozen-in
properties, and the fluid motions can distort any ${\bf V}$ in
a complex fashion. In particular, when the
dynamical system
$$
\dot{\bf x} = {\bf u}
\eqno(3)
$$
produces chaotic motion, the effect on the distribution of the
advective fields can be rich and surprising, giving rise to
intense local concentrations and lacunae of fine structure.

In real fluid settings the situation may be more complicated
than even this description suggests.  If either $S$ or ${\bf V}$
can feed back on the dynamics of ${\bf u}$ itself, the
equations lose even their superficially linear appearances.
For ordinary fluid motions, we have $\rho{\bf V}=
{\pmb{\omega}}$,  where $\rho$ is the fluid density and the
problem is no longer kinematic. Rather, ${\bf
u}$ satisfies
$$
\rho(\partial_t {\bf u} + {\pmb{\omega}}{\pmb{\times}}{\bf u}) =  -
{\bf {\nabla}}p
  -{1\over 2}\rho {\bf {\nabla}}{\bf u}^2
+ {\bf F} + {\rm thermal\ viscous\ effects},
\eqno(4)
$$
where ${\bf F}$ is an external force density.  So when $\rho{\bf V}$ is
vorticity, we have coupled equations for ${\bf u}$ and ${\bf V}$.

G.I. Taylor, who made early contributions to the study of
(1) for turbulent ${\bf u}$, observed that vorticity is concentrated
by turbulence.$^2$  To learn what properties of the motion favor this
effect, we may begin with the study of (2) without worrying, at first,
whether these motions correspond to solutions of
(4).  That leads to the search for a field ${\bf u}$ that may
produce local vorticity enhancement when introduced into (2).
We call this the {\sl kinematic} turbulence problem, after the
usage of dynamo theory, where (2) applies when $\rho {\bf V}$ is
the magnetic field.

Batchelor and others have sought analogies between the vorticity and
magnetic field, since they are both controlled by equation (2).
However, this viewpoint has been belittled
because ${\pmb{\omega}}={\pmb{\nabla}}{\bf\times u}$, while no
analogous relation exists  between ${\bf B}$ and $u$.
Yet this relation is implied by
(2) and (4), and it need not as a result be considered explicitly.
When ${\bf B}$ is the field in question, we
couple it to (4) through a nonlinear term representing the Lorentz force.
That is,
$$
\rho(\partial_t {\bf u} + {\bf u \cdot \nabla u}) =
-{\bf \nabla}p - {1\over 4\pi} {\bf B \times \nabla\times B}
+ {\bf F} + {\rm thermal\ viscous\ effects}. \eqno(5)
$$
Typically the kinematic
aspects of the dynamo problem are stressed over those
of the analogous vorticity problem.  But if the
analogy is good, it is not clear whether this should
raise interest in the kinematic turbulence problem or
throw the kinematic dynamo problem in a bad light.
In any case, there does seem to be considerable
interest in studying the effect of fluid motion on an immersed vector
field through (2), for prescribed ${\bf u}$.

These general thoughts lead into our discussion
of the effect of a chaotic motion on a
vector field satisfying (2).
We describe in general terms the procedures that we
have been following,
omit the most strenuous portions of the calculations and refer to
the related work on the mixing of a passive scalar by a
continuous flow$^3$,
and to the study by Aurell and Gilbert$^4$ of fast dynamos in discrete
maps.

\bigskip
\goodbreak
\centerline{\bf THE FORMAL SOLUTION}
\medskip

We consider velocity fields
$\bf u$ that are steady.  In this autonomous case, the
problem can be simplified by writing ${\bf V}({\bf x},t) = {\bf
V}_0({\bf x})\exp(\lambda t)$.  Then we obtain, for the diffusionless
case,
$$
{\cal D} {\bf V} / {\cal D}t \equiv
({\bf u\cdot\nabla}{\bf V}_0 - {\bf V}_0{\bf \cdot\nabla u})
e^{\lambda t} =
-\lambda{\bf V}_0e^{\lambda t}.
\eqno(6)
$$
We may look for a solution of the form ${\bf V}_0 =
q({\bf x}){\bf u}({\bf x})$, where
$$
{\bf u\cdot\nabla}q = -\lambda q.\eqno(7)
$$
This shows how the kinematic problem for vector fields may be
related to the more extensively studied problem of passive scalar
transport$^3$; if we set $S({\bf x},t)=q({\bf x})\exp(\lambda t)$
in (1), we get (7).  Moreover, if $q$ is constant on streamlines,
then we must
have $\lambda=0$.  These special solutions arise in the very
restricted conditions of steady flow without dispersive effects,
and they illustrate the kind of degeneracy that we encounter when the
conditions are too simple.

More generally, we can write the solution of (2) formally, as
shown by Cauchy.
Let ${\bf x}(t,{\bf a})$ be the position of the fluid particle
that was at the point
${\bf a}$ when $t=0$.  Then the field evolves according to
$$
{\bf V}({\bf x},t) = {\cal J}({\bf a},t) {\bf V}({\bf a},0)
\quad,
\eqno(8)
$$
where ${\cal J}({\bf a},t)=\partial({\bf x})/\partial({\bf a})$
is the Jacobian matrix of the transformation that moves the fluid into
itself with time with ${\bf x}={\bf x}({\bf a},t)$.

We write ${\bf x}={\pmb{\varphi}}\,^t{\bf a}$, where
${\pmb{\varphi}}\,^t$ is the flow that maps the initial positions
of the fluid particles into their positions at time $t$.
Its inverse, ${\bf a}={\pmb{\varphi}}\,^{-t} {\bf x}$, maps
particles at time $t$ and position $\bf x$ back to their initial
positions.
Then
we can write (8) in the seemingly complicated, but quite useful form,
$$
{\bf V}({\bf x},t) = \int \delta({\bf a}-{\pmb{\varphi}}^{-t}{\bf x})
{\cal J}({\bf a},t)\,{\bf V}({\bf a},0)\, d^3{\bf a}
\quad,
\eqno(9)
$$
where the integral operator introduced here is the analogue of the
Perron-Frobenius operator for the case of scalar advection.$^5$
Having turned the differential
equation (2) into an integral equation, we may use analogues
of Fredholm's methods$^6$ to solve it.

If we were to include the effects of diffusion by using a term
of the form  $\eta\nabla^2{\bf V}$ on the  right hand side of
(2), with $\eta$ constant, we would need to replace
${\cal D}/{\cal D}t$ by ${\cal D}/{\cal D}t -
\eta\nabla^2$ in (6) and modify (9),
but for now we shall speak only of the noise-free case.

\bigskip
\goodbreak
\centerline{\bf STRUCTURES OF CHAOTIC FLOWS}
\medskip

To describe the evolution of a frozen-in magnetic field, we need
a suitable means to characterize the flow.  For the chaotic flows
on which we concentrate here, the Lagrangian orbits determined by (3)
densely contain unstable periodic orbits.  Their union forms what we
may call a chaotic set.  If object is attracting, the term
{\it strange attractor} is appropriate while in the opposite case it
becomes a strange {\it repeller,} from which representative points depart
as time proceeds. Even though all periodic orbits within such an
ensemble are unstable, trajectories of the flow often spend extended
periods tracking these paths.
The strange set provides a delicate, skeletal structure
embedded within the flow that can be used to systematically
approximate such properties
as its local stability and geometry.$^7$
This is the basic idea underlying
{\sl cycle expansions}$^{8}$, the method that we apply here to the problem of
vector advection.

To be explicit, we
consider a trajectory $x(t)$ generated by the ordinary differential
equation
$$
\r {\ddd}x +\dot x - cx+x^3 = 0
\quad,
\eqno(10)
$$
with parameter $c$. This  is a special case of equations arising
in multiply diffusive convection$^{9}$. The absence of a second derivative
in (10) ensures that the flow is solenoidal, in line with the
kinds of flow most
commonly studied in dynamo theory, in spite of the
fact that every putative dynamo is compressible.
For certain values of $c$, homoclinic orbits exist and we let $c$
be close to such a value and, in particular, one such that in the
neighborhood of the origin the flow
satisfies Shil'nikov's criterion for the
existence of an infinity of unstable periodic orbits.$^{10}$
Then, an infinite sequence of intertwined
saddle-node and period-doubling bifurcations creates a dense chaotic set,
and the motion of points in the flow is chaotic.

Details of the structure of this flow are presented elsewhere;$^{11}$
here we mention only that $x(t)$, under the conditions mentioned, is a
sequence of pulses.  Moreover, the separation of the $k${-th} and
$(k-1)${-st} pulses, $\Delta_k$, may be expressed in terms of the
two previous spacings.  This provides us with a {\it timing map},
which is a two-dimensional map resembling in form the H\'enon map$^{12}$.
Thus,
to good approximation, $$
\Delta_k = \bar T + \alpha \tau_k
\quad,
\eqno(11)
$$
where $\bar T$ is a mean period, $\alpha$ is a small parameter,
and $\tau_k$ an irregular timing fluctuation satisfying,
$$
\tau_{k+1} = 1-a\tau_k^2 - \tau_{k-1}
\quad,
\eqno(12)
$$
which is the orientation and area preserving form of the H\'enon map.

The form of any particular pulse is quite close to that of the
homoclinic orbit, which can be computed at the outset where
the map (12) determines a sequence of pulse positions. Once these
are known, we can generate a complete and reasonably accurate solution
for the velocity field.  Thus we can reconstruct the entire {\em flow}
from this simple map.  Moreover, this map also contains the invariant
information
about the periodic orbits of the chaotic set (cycle topology and
stability eigenvalues) and serves as a powerful tool in
the construction of cycle expansions.

In the following sections, we describe the technique of cycle expansions
for the problem at hand. This needs no explicit specification of the
velocity field $\bf u$, but in the concluding section we report
some numerical results obtained for the particular flow modeled by (12).

\bigskip
\goodbreak
\centerline {\bf EVOLUTION AND TRANSFER OPERATORS}
\medskip

A frozen-in vector field is stretched, squeezed and
swept around by the chaotic flow, evolving as described by (8).
To compute the large-scale evolution of the field,
we rewrite (9) as
$$
V_i(\bfx,t) = \int_\Sigma d^3a \; \cll^t_{ij} (\bfx,\bfa) V_j(\bfa,0)
\quad,
\eqno(13)
$$
with a kernel
$$
\cll_{ij}^t(\bfx,\bfa) =
\delta(\bfa-\bfp^{-t}\bfx){ \pde x_i \over \pde a_j}
\quad,
\eqno(14)
$$
where summation over repeated indices is understood.
The kernel $\cll^t_{ij}$ controls the evolution of the embedded vector
field. This {\it transfer operator} is linear and possesses (for nice
hyperbolic systems, the so-called Axiom A flows) a
sequence of eigenvalues, $e^{-\nu_0 t},e^{-\nu_1 t},e^{-\nu_2 t},...\;.$
For large times, the effect of $\cll^t$ is dominated by its leading
eigenvalue, $e^{-\nu_0 t}$ with $Re(\nu_0) < Re(\nu_i)$, $i=1,2,3,...$.
In this way the transfer operator furnishes the fast dynamo rate,
$\nu\equiv-\nu_0$.

The operator $\cll_{ij}^t$ was introduced in ref.~8 in order to study the
``stability of strange sets'', and applied to discrete map
models of fast dynamos in ref.~4. Here we apply it to continuous
time flows.

\bigskip
\goodbreak
\centerline{\bf A TRACE FORMULA}
\medskip

To calculate the leading eigenvalue of $\cll^t$, we
evaluate the trace
$$
\tr (\cll^t) = \int_\Sigma d^3 a \; \cll_{ii}^t (\bfa,\bfa)
=\int_{\Sigma} d^3 a \; \delta(\bfa-\bfp^{-t}\bfa)
{\partial \varphi_i \over \partial a_i}
\quad,
\eqno(15)
$$
which asymptotes to $e^{\nu t}$ for long times.
We evaluate this integral by means of explicit
periodic orbit expansions. Each cycle within the fabric of the flow
contributes to the overall trace of the operator, and
each contribution is obtained by integration along the periodic
orbit in suitable local coordinates.  We omit the details of
this purely technical operation (see ref.~3) and report that
each prime cycle $p$ together with its repeats contributes a term
$$
T_p
\sum_{r=1}^\infty
{\tr \; {\bf J}_p^r\over | \det ({\bf 1} - {\bf J}_p^{-r})|}
\delta(t-rT_p)
 \quad
 \eqno(16)
$$
to the integral. $T_p$ is the cycle period and
${\bf J}_p$ is the  transverse
stability matrix
$\hat{{\bf u}}(t+T_p)= {\bf J}_p \hat{{\bf u}}(t)$ for a two-vector in
the tangent plane transverse to the flow.
The ${\bf J}_p$ eigenvalues $\Lambda_{p,1}$, $\Lambda_{p,2}$
are independent of the position along the orbit and the choice of transverse
coordinates.

The trace of the transfer operator is the sum over all
periodic orbit contributions, with each cycle weighted by
its intrinsic stability
$$
\tr({\cal L}^t)=\sum_p T_p \sum_{r=1}^{\infty}
{\tr \; {\bf J}_p^r\over | \det ({\bf 1} - {\bf J}_p^{-r})|}
\delta(t-rT_p)
 \eqno(17)
$$
$$
={i\over 2\pi} \int_0^\infty dk \; e^{ikt} {\partial\over\partial k}
\sum_p\sum_r {1\over r}
{\tr \; {\bf J}_p^r \over |\det ({\bf 1} - {\bf J}_p^{-r}|}
 e^{-ikrT_p}
 \quad,
 \eqno(18)
$$
where we have introduced the Fourier representation of Dirac delta
functions.

\bigskip
\goodbreak
\centerline {\bf A FREDHOLM DETERMINANT}
\medskip

We rotate the axes  by the Wick transformation, $k \rightarrow is$,
and observe that the summations in (18) correspond to
the logarithmic derivative
of the function
$$
F(s) = {\rm exp}  \left[ -
         \sum_{p} \sum_{r=1}^\infty {1 \over r}
 {\tr \ {\bf J}_p^{r}\over |\det ({\bf 1}-{\bf J}_p^{-r})|}
  e^{srT_p}
                 \right]
\quad,
\eqno(19)
$$
related to the above trace by
$$
\tr (\cll^t) = {i\over 2 \pi} \int_{-i\infty}^{i\infty} ds
\,\, e^{-st} {F'(s) }/{ F(s)}
\quad.
\eqno(20)
$$

The function $F(s)$ is a {\sl Fredholm determinant}. Its zeros
produce singularities in the integrand of (20). If it is an {\sl
entire function} (defined over the whole complex plane), then the
contour of integration can be suitably deformed so as to encircle the
various poles of $F'(s)/F(s)$, and the trace becomes a sum over the
residues of the integrand:
$$
\tr (\cll^t) =
\sum_{n=0}^\infty m_n e^{- \nu_n t}
\quad,
\eqno(21)
$$
where $\nu_n$ is a pole of multiplicity $m_n$.
Hence, the spectrum of the transfer operator
is determined by the zeros of the Fredholm determinant $F(s)$.

In order to simplify $F(s)$,
we factor the denominator cycle stability determinants
into products of expanding and contracting eigenvalues.
The example at hand is
a 3-dimensional hyperbolic flow with cycles possessing
one expanding eigenvalue $\Lambda_p$ (of
absolute value $>1$), and one contracting eigenvalue $\lambda_p$, with
$|\lambda_p|<1$.
Then the determinant may be expanded as follows:
$$
{|\det \left( {\bf 1}-{\bf J}_p^{-r} \right)|}^{-1}
 =  | (1-\Lambda_p^{-r}) (1-\lambda_p^{-r}) | ^{-1}
 =   |\lambda_p|^{r} \sum_{j=0}^\infty \sum_{k=0}^\infty
   \Lambda_p^{-jr} \lambda_p^{kr}
\quad.
\eqno(22)
$$
With this decomposition
we can rewrite the exponent in (19) as
$$
    \sum_p \sum_{r=1}^\infty {1 \over r}
      { (\lambda_p^r+\Lambda_p^r)e^{s rT_p}
         \over |\det \left( {\bf 1}-{\bf J}_p^{-r} \right)| }
 =  \sum_p \sum_{j,k=0}^{\infty}   \sum_{r=1}^\infty
{1 \over r}
\left( |\lambda_p| \Lambda_p^{-j} \lambda_p^k e^{sT_p}  \right)^r
                (\lambda_p^r+\Lambda_p^r)
\quad,
\eqno(23)
$$
which has the form of the expansion of a logarithm:
$$
\sum_p \sum_{j,k}
\left[
\log \left(1 - e^{sT_p} |\lambda_p| \Lambda_p^{1-j} \lambda_p^{k}\right) +
\log \left(1 - e^{sT_p} |\lambda_p| \Lambda_p^{-j} \lambda_p^{1+k}\right)
\right]
\quad.
\eqno(24)
$$
The Fredholm determinant is therefore of the form,
$$
F(s) = F_e(s) F_c(s)
\quad,
\eqno(25)
$$
where
$$
F_e(s) =  \prod_p  \prod_{j, k=0}^\infty
\left( 1 - t_p^{(jk)} \Lambda_p \right)
\quad,
\eqno(26)
$$

$$
F_c(s) =  \prod_p  \prod_{j, k=0}^\infty
\left( 1 - t_p^{(jk)} \lambda_p \right),
\eqno(27)
$$
with
$$
t_p^{(jk)} = e^{sT_p} |\lambda_p| {\lambda_p^k \over \Lambda_p^j}
\quad.
\eqno(29)
$$
The two factors present in $F(s)$ correspond to the two Floquet
exponents of the periodic orbits, the expanding and contracting exponents.
In form, they are the Selberg-type
products known in other areas of mathematical physics.$^{13}$

\bigskip
\goodbreak
\centerline{\bf DYNAMOS FOR MAPS VS. FLOWS}
\medskip

For 2-$d$ Hamiltonian volume preserving systems
$\lambda = 1/\Lambda$, and (26) reduces to
$$
F_e(s) =  \prod_p \prod_{k=0}^\infty
   \left( 1 - { t_p \over \Lambda_p^{k-1} } \right)^{k+1},
\quad t_p=\frac{e^{sT_p}}{\mid \Lambda_p \mid} \quad.
\eqno(30)
$$
Denoting the eigenvalue sign by $\sigma_p =  \Lambda_p/| \Lambda_p|$,
the Hamiltonian zeta function (the $j=k=0$ part of the product (27))
is given by
$$
1/\zeta_{dyn}(s) =  \prod_p  \left( 1 -   \sigma_p e^{ s T_p } \right)
\, .
\eqno(31)
$$
This is a curious formula: the Hamiltonian
zeta function depends only on the return times, not on the
eigenvalues of the cycles.
Furthermore, the identity
$$
{\Lambda + 1/\Lambda
        \over
 |(1-\Lambda)(1-1/\Lambda)|}
         = \sigma + {2 \over |(1-\Lambda)(1-1/\Lambda)|}
$$
substituted into (25) leads to a relation between the vector
and scalar advection Fredholm determinants:
$$
F_{dyn}(s)=F_0(s)^2/\zeta_{dyn}(s)
\,.
\eqno(32)
$$
The Fredholm determinants in this equation are entire for nice
hyperbolic
(axiom A) systems, since  both of them correspond to
multiplicative operators$^{14,15}$. For {\em maps} with finite Markov
partition the inverse zeta function
(31) reduces to a polynomial since the time
$T_p=n_p$ is an integer, and the curvature terms$^{8}$
in the cycle expansion vanish. For example, for maps
with complete binary partition, and with the fixed point stabilities
of opposite signs, the cycle expansion reduces
to
$$
1/\zeta_{dyn}(s)=1.
\eqno(33)
$$
For such {\em maps} the dynamo Fredholm determinant is simply the square
of the scalar advection Fredholm determinant, and therefore all its zeros
are double. In other words, for maps, the fast dynamo
rate equals to the scalar advection rate$^{16}$.

However, for {\em flows} the dynamo effect is distinct from the scalar
advection. For example, for flows with finite symbolic dynamics grammar,
(32) implies that the dynamo zeta function
is a ratio of two entire determinants:
$$
1/\zeta_{dyn}(s)=F_{dyn}(s)/F_{0}^2(s)
\,.
\eqno(34)
$$
This relation implies that for {\em flows}
the zeta function has double poles at
the zeros of the scalar advection Fredholm determinant, with zeros of
the dynamo Fredholm determinant no longer coinciding with the zeros of the
scalar advection Fredholm determinant;
the leading zero of the dynamo Fredholm determinant is larger than the
scalar advection rate, and the rate of decay of the magnetic field is no
longer governed by the scalar advection.

\bigskip
\goodbreak
\centerline{\bf NUMERICAL INVESTIGATIONS}
\medskip

For our numerical investigations of the kinematic dynamo effect
we have used (11)
with ${\overline{T}} =1$, $a=6$ and $\alpha\in [0,1]$.
For $a=6$ the H\'enon map has a
repeller with complete binary grammar.
We have computed all periodic orbits up to length 13.
The period of the prime cycle $p$ is obtained
by summing up the contributions (11):
$$
T_p=n_p+\alpha\sum_{i=0}^{n_p-1}x_{p,i}
\, .
\eqno(35)
$$
where $x_{p,i}$ are the periodic points in the prime cycle $p$.
In table 1 we list the stabilities
and $\sum x_{p,i}$ for the cycles  up to length $n_p\leq 6$.
 From such listing one can generate the prime periods for any
parameter value $\alpha$, feed this data into cycle expansions,
and extract a set of leading eigenvalues by determining
the zeros of $F(s)$ in the complex plane by standard
root finding routines such as the
Newton-Raphson method. A visualization of
the complex function $F(s)$ is afforded  by a contour plot
of $\ln |F(s)|$. In such plot the eigenvalues correspond to
the minima of $\ln |F(s)|$, and the range of validity of a
cycle expansion is indicated by absence of fine structure beyond
some Re($s$).

Figs. 1(a) and 1(b) show the
contour plot of  Fredholm determinants for the scalar and vector advection
respectively, for $\alpha=0$.
For this case (constant return time) the dynamics is equivalent to a map,
and the dynamo Fredholm determinant
should  have double zeros coinciding with the zeros of the scalar
advection Fredholm determinant. In this computation we used all prime cycles up
to topological length $12$, with polynomial truncation
yielding the $12$ leading zeros in both cases. Since the zeros of
the dynamo Fredholm determinant are double zeros,
in the polynomial truncation only the
first few are accurate, with the nonleading terms
converging more poorly.

The result for maps should be contrasted to the result for flows:
figs. 2(a) and 2(b) show the corresponding
Fredholm determinants for a flow with large dispersion of
return times, $\alpha=1$.
Here the zeros of the two Fredholm determinants do not
coincide, and the leading zero of the dynamo Fredholm determinant
yields the dynamo rate. Parenthetically, this lifting of eigenvalue
degeneracy is very noticeable already for return time dispersion as small
as $\alpha=0.01$.
Fig. 3 shows the contour plot corresponding to the dynamo zeta
function (31). The leading zeros
coincide with the leading zero of the dynamo Fredholm determinant,
but this zeta has a double pole at the escape rate zero
of $F_0(s)$, see eq. (34), which is very apparent in this plot.
In contrast, the zeta function (33) for a discrete time map is trivial.

Note that the dynamo zeta function does not require evaluation
of cycle eigenvalues $\Lambda_p$,
and even so the pole in the zeta yields the escape rate;
we can compute the escape rate from the cycle periods $T_p$
only by
locating the first pole of the dynamo zeta function.
This indicates an intimate and not yet elucidated
connection between
the periods and stabilities of general dynamical systems.

Another interesting property of the fast dynamo flows is
that in these systems
the magnetic field can grow exponentially although the flow
is repelling, the escape rate is positive, and scalar advection
density decreases exponentially.
This is illustrated by the difference in the sign of the
leading eigenvalues in figs. 2(a) and 2(b).

\bigskip
\goodbreak
\centerline{\bf CONCLUSIONS}
\medskip

We have introduced a new transfer operator for chaotic flows whose leading
eigenvalue yields the dynamo rate of the fast kinematic dynamo
and applied cycle expansion
of the Fredholm determinant of the new operator to evaluation
of its spectrum.
The theory has been tested on a normal form model of the
vector advecting dynamical flow. If the model is a simple map with
constant time between two iterations, the dynamo rate is
the same as the escape rate of scalar quantities.
However, a spread in Poincar\'e section return times
lifts the degeneracy of the vector and scalar advection rates,
and leads to dynamo rates that dominate over
the scalar advection rates. For sufficiently large
time spreads we have even found repellers for which
the magnetic field {\em grows} exponentially,
even though the scalar densities are decaying exponentially.

\bigskip
\goodbreak
\centerline{\bf ACKNOWLEDGEMENTS}
\medskip

This work
has been supported at Columbia University by the A.F.O.S.R. under grant no.
AFOSR89-0012. N.J.B. thanks the S.E.R.C. for a postdoctoral fellowship.
P.C. thanks the Carlsberg Foundation for the support, and
M.J.~Feigenbaum
for the hospitality at the Rockefeller University, where part
of this work was done. G.V. thanks the Sz\'echenyi Foundation
and OTKA grant F4286 for the support, and the Chaos and Turbulence
Studies Center, Niels Bohr Institute, for the hospitality.
P.C. thanks to E.~Aurell for communicating V.~Oseledec's results.

\bigskip
\goodbreak
\centerline{\bf REFERENCES}
\medskip

\begin{enumerate}
\item Ottino, J.M., ``The kinematics of mixing: stretching,
chaos and transport'' (Cambridge, 1989).

\item Taylor, G.I. 1938. Proc. R. Soc. London {\bf A164:} 15.

\item Cvitanovi\'c, P. and Eckhardt, B.  1991. J. Phys. A {\bf 24:}
L237.

\item Aurell, E. and Gilbert, A. 1993. Geophys. Astrophys. Fluid
Dynamics, in press.

\item Ruelle, D.
1989.
Commun. Math. Phys. {\bf 125:} 239.

\item 
Courant, R. and Hilbert, D. 1953. Methods of Mathematical Physics,
Volume 1 (Interscience publishers);
A. Grothendieck, A. 1956.
        Bull. Soc. Math.  France {\bf 84:} 319. 

\item Cvitanovi\'c, P. 1991.
          Physica {\bf D 51:} 138.

\item Artuso, R., Aurell, E. and Cvitanovi\'c, P. 1990.
Nonlinearity {\bf 3:} 325 and 361.

\item Arneodo, A., Coullet, P. and Spiegel, E.A. 1985
Geophys. Astrophys. Fluid Dynamics {\bf 31:} 1.

\item Shil'nikov, L.P. 1965. Soc. Math. Dokl. {\bf 6:} 163.
Shil'nikov, . 1970. Math. USSR Sbornik {\bf 10:} 91.
Tresser, C. 1984. Ann. Inst. H. Poincar\'e {\bf 40:} 441.

\item Balmforth, N.J., Ierley, G.R. and Spiegel, E.A. 1993.
Submitted to SIAM J. Applied Math.

\item H\'enon, M. 1976. Commun. Math. Phys. {\bf 50:} 69.

\item Selberg, A. 1956. J. Indian Math. Soc. {\bf 20:} 47.

\item Ruelle, D. 1986.  J. Stat. Phys. {\bf 44:} 281.

\item Rugh, H.H. 1992. Nonlinearity {\bf 5:} 1237.

\item Oseledec, V. (private communication to E. Aurell).
\end{enumerate}

%
%
%
%
%
%
%
\bigskip
\goodbreak
\centerline{\bf FIGURE CAPTIONS}
\medskip

{\bf Figure 1:}  Contour plot of $\log | F(s) |$ for
({a})
the scalar advection Fredholm determinant,  and for
({b})
the dynamo Fredholm determinant
for the H\'enon map $\alpha=0$,
using prime cycles up to topological length $12$.
Note that the scalar and vector advection rates coincide.

\medskip
{\bf Figure 2:}  Contour plot of $\log | F(s) |$ for
({a})
the scalar advection Fredholm determinant for the flow (35), and
({b})
the dynamo Fredholm determinant for the
$\alpha=1$ flow,  using prime cycles up to topological length $12$.
Note that for flows the vector advection eigenvalue (fast dynamo rate)
dominates the scalar advection eigenvalue.

\medskip
{\bf Figure 3:} Same as fig. 2 for the dynamo zeta.
The leading eigenvalue yields the fast dynamo rate $\nu$,
and the convergence radius of the cycle expansion is controlled by the
pole at the scalar advection eigenvalue eq. (34).

\bigskip
\bigskip

\begin{tabular}{|c|r|r|l|}
\hline
period & $\Lambda_p$~~~~~~~~~~~~
                            &    $\sum x_{p,i}$~~~~~~~~~~~~
                                                 & code\\ \hline
1 & 0.715167524380$\times 10^1$ & -0.60762521851077 & 0 \\
1 &-0.295284632592$\times 10^1$ &  0.27429188517743 & 1 \\
   \hline
2 &-0.989897948557$\times 10^1$ &  0.33333333333333 & 10 \\
   \hline
3 &-0.131907273972$\times 10^3$ & -0.20601132958330 & 100 \\
3 & 0.558969649960$\times 10^2$ &  0.53934466291663 & 110 \\
   \hline
4 &-0.104430107304$\times 10^4$ & -0.81649658092773 & 1000 \\
4 & 0.577998269891$\times 10^4$ &  0.00000000000000 & 1100 \\
4 &-0.103688325098$\times 10^3$ &  0.81649658092773 & 1110 \\
   \hline
5 &-0.760653437184$\times 10^4$ & -1.42603220657928 & 10000 \\
5 & 0.444552400077$\times 10^4$ & -0.60665407777388 & 11000 \\
5 & 0.770202485970$\times 10^3$ &  0.15137550164056 & 10100 \\
5 &-0.710688356166$\times 10^3$ &  0.24846322760447 & 11100 \\
5 &-0.589498852840$\times 10^3$ &  0.87069547289495 & 11010 \\
5 & 0.390994248124$\times 10^3$ &  1.09548541554650 & 11110 \\
   \hline
6 &-0.545745270604$\times 10^5$ & -2.03413425566653 & 100000 \\
6 & 0.322220609858$\times 10^5$ & -1.21525043702153 & 110000 \\
6 & 0.513761651093$\times 10^4$ & -0.45066243593297 & 101000 \\
6 &-0.478461466317$\times 10^4$ & -0.36602540378444 & 111000 \\
6 &-0.639399984360$\times 10^4$ &  0.33333333333333 & 110100 \\
6 &-0.639399984360$\times 10^4$ &  0.33333333333333 & 101100 \\
6 & 0.390193872690$\times 10^4$ &  0.54858377035486 & 111100 \\
6 & 0.109490945979$\times 10^4$ &  1.15146335826616 & 111010 \\
6 &-0.104338416941$\times 10^4$ &  1.36602540378444 & 111110 \\
   \hline
\end{tabular}

\vskip 5mm

Table 1.
All periodic orbits up to 6 bounces
for the Hamiltonian H\'enon mapping (11), $a=6$.
The columns list the topological length of the cycle,
its expanding eigenvalue $\Lambda_p$,
the period of the orbit, and the binary code for the cycle.

\end{document}